\documentclass[reprint,superscriptaddress,amsmath,amssymb,aps,prx,longbibliography]{revtex4-1}

\usepackage{graphicx}
\usepackage{dcolumn}
\usepackage{bm}
\usepackage[colorlinks=true,urlcolor=blue,linkcolor=blue,citecolor=blue]{hyperref}
\usepackage[normalem]{ulem}
\usepackage{amsmath}
\usepackage{amssymb}
\usepackage{bbold}

\usepackage[english]{babel}
\usepackage[dvipsnames]{xcolor} 
\usepackage{float}
\usepackage{comment}
\DeclareUnicodeCharacter{2009}{\,} 

\newcommand{\mat}[1]{\mathsf{#1}}

\begin{document}

\title{Ergodic inclusions in many body localized systems}

\author{Luis Colmenarez}
\email{colmenarez@physik.rwth-aachen.de}
\affiliation{Institute for Quantum Information, RWTH Aachen University, 52056 Aachen, Germany}
\affiliation{Institute for Theoretical Nanoelectronics (PGI-2), Forschungszentrum Jülich, 52428 Jülich, Germany}
\affiliation{Max Planck Institute for the Physics of Complex Systems, N\"othnitzer Stra{\ss}e~38, 01187-Dresden, Germany}
\author{David J. Luitz}
\email{dluitz@uni-bonn.de}
\affiliation{Institute of Physics, University of Bonn, Nußallee 12, 53115 Bonn, Germany}
\author{Wojciech De Roeck}
\email{wojciech.deroeck@kuleuven.be}
\affiliation{KU Leuven, 3001 Leuven, Belgium}
\date{\today}

\date{\today}

\begin{abstract} 

We investigate the effect of ergodic inclusions in putative many-body localized systems. We consider the random field Heisenberg chain, which is many-body localized at strong disorder and we couple it to an ergodic bubble, modeled by a random matrix Hamiltonian. Recent theoretical work suggests that localized systems are unstable to ergodic bubbles, driving the delocalization transition. We tentatively confirm this by numerically analyzing the response of the on-site purities to the insertion of the bubble. For a range of intermediate disorder strengths, this response decays very slowly, or not at all, with increasing distance to the bubble. This suggests that at those disorder strengths, the system is delocalized in the thermodynamic limit. However, artefacts in the numerics cannot be ruled out.
 
\end{abstract} 

\maketitle

\section{Introduction}
The discovery that noninteracting particles in a disorder potential can become completely immobile by P. W. Anderson in 1958 \cite{anderson_absence_1958} has created a new field of study, which became enormously active in the last decade after it was predicted that this localized phase could persist in the presence of interactions, leading to a \emph{perfect insulator} at any temperature \cite{fleishman_interactions_1980,gornyi_interacting_2005,basko_metalinsulator_2006,oganesyan_localization_2007,imbrie_many-body_2016,imbrie_diagonalization_2016}. While it is difficult to realize such systems in condensed matter experiments due to the presence of phonons, many-body localized (MBL) systems were realized in synthetic quantum matter \cite{schreiber_observation_2015,luschen_signatures_2017,smith_many-body_2016}.
It was first suggested that the interplay of interaction and disorder gives rise to a non-equlibrium phase transition between a thermal phase at weak disorder, which satisfies the eigenstate thermalization hypothesis (ETH) \cite{deutsch_quantum_1991,srednicki_chaos_1994,rigol_thermalization_2008,dalessio_quantum_2016} and a many-body localized phase at strong disorder. Such a transition would be unparalleled in equilibrium \cite{nandkishore_many-body_2015,vosk_theory_2015,abanin_recent_2017,imbrie_local_2017,luitz_ergodic_2017,agarwal_rare-region_2017,alet_many-body_2018,abanin_colloquium_2019}. A large body of theoretical work now supports the picture that the many-body localized phase is characterized by an emergent complete set of quasi-local integrals of motion \cite{serbyn_local_2013,huse_phenomenology_2014}, which are fully consistent with the observed phenomenology of area law entanglement in all many-body eigenstates \cite{bauer_area_2013,luitz_long_2016,yu_bimodal_2016} as well as with the logarithmic post-quench entanglement production \cite{chiara_entanglement_2006,znidaric_many-body_2008,bardarson_unbounded_2012,serbyn_universal_2013}.


Recently, however, the stability of the insulating phase in the thermodynamic limit has been put into question, generating a debate on MBL as a phase of matter \cite{abanin_distinguishing_2021,panda_can_2020,ghosh_resonance-induced_2022,kiefer-emmanouilidis_slow_2021,luitz_absence_2020,morningstar_avalanches_2022,sels_bath-induced_2022,sels_dynamical_2021,sels_thermalization_2023,sierant_thouless_2020,suntajs_ergodicity_2020,suntajs_quantum_2020,sierant_polynomially_2020,sierant_challenges_2022}. Some works suggest that the critical disorder might be way higher than what was expected \cite{morningstar_avalanches_2022,sels_bath-induced_2022,weiner_slow_2019,crowley_constructive_2022,doggen_many-body_2018}, others even predict an infinite critical disorder in the thermodynamic limit \cite{sels_dynamical_2021,sels_thermalization_2023,kiefer-emmanouilidis_slow_2021,suntajs_quantum_2020,evers_internal_2023}. As a matter of fact, quantum simulations have shown MBL signatures in systems with few tens of particles \cite{schreiber_observation_2015,smith_many-body_2016,kondov_disorder-induced_2015,gong_experimental_2021}, thus today's experiments can only access a MBL regime that asymptotically, in time and system size, may or may not thermalize. 
One central aspect in this debate is the delocalization transition mechanism and the crossover behavior in finite systems, whose understanding is still incomplete.
In the vicinity of the transition, anomalously slow dynamics was observed \cite{bar_lev_absence_2015,agarwal_anomalous_2015,luitz_extended_2016,luschen_observation_2017,lezama_apparent_2019}, which was related to anomalous thermalization behavior \cite{luitz_long_2016,luitz_anomalous_2016,roy_anomalous_2018,colmenarez_statistics_2019,znidaric_diffusive_2016} and a theoretical description based on rare insulating inclusions was proposed \cite{gopalakrishnan_griffiths_2016,potter_universal_2015,vosk_theory_2015}. It remains however unclear how this picture can be reconciled with the observation of slow dynamics in quasiperiodic potentials \cite{luschen_observation_2017,lev_transport_2017,luitz_ergodic_2017}. It was also proposed that the many-body resonances are driving the slow dynamics in this regime \cite{long_phenomenology_2022}.

\begin{figure}[t] 
	\centering
	\includegraphics[width=8.5 cm]{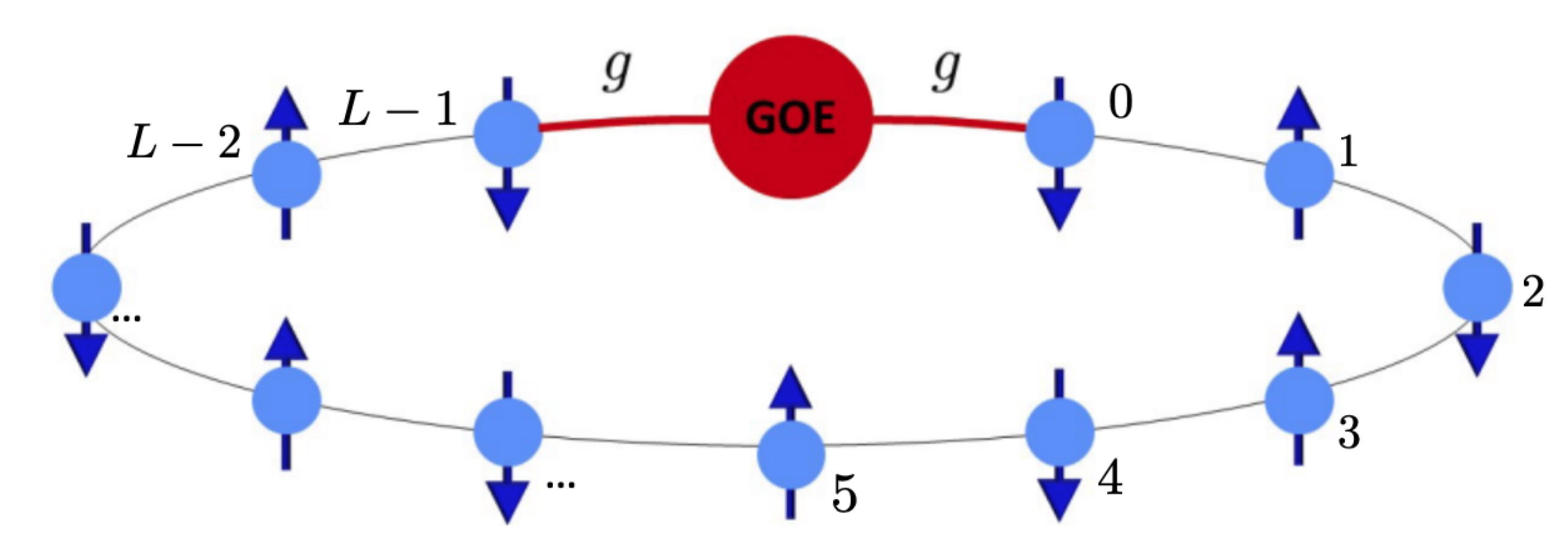}
	\caption{System setup: A spin chain (blue) is coupled to an ergodic bubble (red). The spins at site 0 and $L-1$ are both coupled to the bubble with coupling strength $g$, such that both "ends" of the chain are symmetric. In order to make the bubble perfectly thermal, its Hamiltonian is given by a random GOE matrix, while the spin chain corresponds to a XXZ spin $1/2$ model in the presence of a random field coupled to local $\hat S_i^z$ operators and nearest neighbour interactions. }  
	\label{fig:setup}
\end{figure} 

It was pointed out in Ref. \cite{de_roeck_stability_2017} that many-body localized systems are unstable under certain conditions towards thermal inclusions by a mechanism dubbed ``avalanche'' \cite{de_roeck_many-body_2017,thiery_many-body_2018}, and such a transition as a function of the localization length was confirmed numerically in idealized models \cite{luitz_how_2017,crowley_avalanche_2020}, also tailored to address implications for the instability of MBL in higher dimensions \cite{de_roeck_stability_2017,ponte_thermal_2017,potirniche_exploration_2019}. Current activities now focus on the identification of this mechanism driving the transition in more realistic models \cite{goihl_exploration_2019,sels_thermalization_2023,de_tomasi_rare_2021, tu_avalanche_2023, tu_localization_2023} and experiments \cite{rubio-abadal_many-body_2019,leonard_probing_2023}. This is challenging and so far direct evidence for the avalanche mechanism in standard MBL models is still lacking. In this work, we directly address the issue of avalanches in a standard MBL model and consider a thermal inclusion coupled to a disordered spin chain.

\section{Model} 

We study the random field XXZ chain, modeled by the Hamiltonian $H_\text{XXZ}$ (``the chain'') coupled to an ergodic (or ``thermal'') bubble  with Hamiltonian
$\hat {\mat R}_0$. The setup for the model is illustrated in Fig.~\ref{fig:setup}. The total Hamiltonian is of the form 
\begin{equation}
    \hat H = \hat H_{\text{XXZ}}\otimes \hat{ \mat 1}  + \hat{ \mat 1} \otimes  \hat{ \mat R }_0 +
    g H_{\text{coupling}}.
    \label{eq:hamiltonian}
\end{equation}
Here the first factor of the tensor product refers to the chain and the second factor refers to the thermal bubble.
The  XXZ Hamiltonian acts on $L$ spins labelled by $i=0,\ldots L-1$; 
\begin{equation}
    \begin{split}
    \hat H_{\text{XXZ}} &= J \sum_{i=0}^{L-2} \left[ \tfrac{1}{2}\left( \hat S_i^+ \hat S_{i+1}^- +
    \hat S_i^- \hat S_{i+1}^+ \right) + \Delta \hat S_{i}^z  \hat S_{i+1}^z \right] \\
    &+ \sum_{i=0}^{L-1} h_i \hat S_i^z.
\end{split}
    \label{eq:hxxz}
\end{equation}
where we have taken the random fields $h_i \in [-W,W]$  drawn from a box distribution. We consider the isotropic point $\Delta=1$. All parameters are measured in $J=1$ units.
The coupling term reads:
\begin{equation}
    \begin{split}
    \hat H_\text{coupling} &=  ( \hat{S}^{z}_{0} \otimes  \hat{ \mat 1 } )\cdot( \hat{ \mat 1} \otimes  \hat{ \mat R }_1 )\cdot( \hat{S}^{z}_{L-1} \otimes  \hat{ \mat 1 } )\\
    &+ [( \hat{S}^{+}_{0} \otimes  \hat{ \mat 1 } )\cdot( \hat{ \mat 1} \otimes  \hat{ \mat R}_2 )\cdot( \hat{S}^{-}_{L-1} \otimes  \hat{ \mat 1 } )+h.c.]
\end{split}
    \label{eq:coupling}
\end{equation}
where the matrices $ \hat{ \mat R }_l, l=0,1,2$ are independent random matrices from the Gaussian Orthogonal Ensemble (GOE), with a scaled bandwidth defined by
\begin{equation}
\mat{R}_{l}= \frac{\beta}{2} \left( \mat{A} + \mat{A}^T \right) \in
    \mathbb{R}^{n_\text{GOE}\times n_\text{GOE}}, \quad \mat{A}_{ij} = \text{norm(0,1)}.
    \label{eq:randommat}
\end{equation}

where $\text{norm(0,1)}$ are normal random variables  with zero mean and unit variance. The dimension $n_{\text{GOE}}$ of the random matrices controls the power of the ergodic bubble, here we use $n_{\text{GOE}}=3,4,5,6,8$. $\beta$ is a real variable that controls the band width of the random matrices and is chosen such that the level mixing (reflected in the overall gap ratio) is maximal for the largest possible range of parameters (see SM Sec.~\ref{sec:beta} for more details). 

Note that the ergodic bubble is coupled to spins $0$ and $L-1$, thus closing the chain into a ring. Therefore, both ends of the chain are symmetric and the longest distance from the bubble corresponds to the spin situated at the middle of the chain $i=\lfloor L/2 \rfloor$. We also note that the coupling in Eq.~\eqref{eq:coupling} of the bubble to the chain is chosen such that all terms preserve the total spin $\hat S_z =\sum_{i=0}^{L-1} \hat S_i^z $ in the  chain. 
This allows us to restrict our study to the largest sector of the Hilbert space, given by $S_z = 0$ for $L$ even and $S_z=1$ for $L$ odd, with a Hilbert space of dimension $\text{dim}\left( \mathcal{H} \right) = n_\text{GOE} \binom{L}{\lfloor L/2\rfloor}$. 
We perform massively parallel shift-invert diagonalization \cite{luitz_many-body_2015,pietracaprina_shift-invert_2018} for computing eigenstates close to the exact spectral center of each sample given by $(E_\text{max}+E_\text{min})/2$, with ($E_\text{min}$)$E_\text{max}$ being the (anti)ground state energy. We confront results for our system with a thermal bubble, Eq.~\eqref{eq:hamiltonian} with the isolated XXZ chain Eq.~\eqref{eq:hxxz}.  \\

\noindent \textbf{Numerical protocol}  We consider two types of observables. 
Firstly, we compute the 
$r_n$-parameter of adjacent energy gaps in the middle of the spectrum given by
$r_n = \text{min} (\delta_{n},\delta_{n+1})/ \text{max} (\delta_{n},\delta_{n+1})$ \cite{oganesyan_localization_2007}, with $\delta_n = E_{n+1}-E_{n}$. Here $E_{n-1},E_n,E_{n+1}$ are consecutive energy levels of the XXZ chain, and the system with a bubble, respectively. This parameter is useful to  distinguish the ergodic and localized phase in a simple way. 

Second, we use the single site purity $\gamma_i = \mathrm{Tr}(\rho^2_i)$, where  $\rho_i = \mathrm{Tr}_{\{L-i\}}(|n\rangle\langle n|)$ is the reduced density matrix on a single site $i$ for a given eigenvector $|n\rangle$.  Due to the $U(1)$ symmetry of the model, $\gamma_i$ can be conveniently expressed via the matrix element $\langle n |S^{z}_i | n  \rangle$ as (a brief analysis of these matrix elements is presented in SM Sec.~\ref{sec:local_magnetization} )
%
\begin{equation}\label{eq:purity_def}
\gamma_i = 2\langle n |S^{z}_i | n  \rangle^2 + \dfrac{1}{2}.
\end{equation}
In an ergodic system, the ETH and random matrix theory predict that $\gamma_i-1/2 \sim d^{-1/2}$ where   $d$ is the Hilbert space dimension, at least for states $|n\rangle$ chosen at maximal entropy \cite{dalessio_quantum_2016}. The matrix elements $\langle n |S^{z}_i | n  \rangle$ by themselves show qualitative signatures of thermalization, see SM Sec.~\ref{sec:local_magnetization} for a short discussion.

In models where ergodicity is induced from boundary effects, as it presumably happens in the case investigated here, one should be more careful and write  $\gamma_i-1/2 \sim  d^{-1/2}_{\text{eff},i}$ where $d_{\text{eff},i}$ is the 
\emph{effective dimension},  see \cite{luitz_long_2016} and SM. 
In contrast, in a localized system, we expect
$\gamma_i$ to depend substantially on the state $|n\rangle$ and the disorder realization, and we expect the average
$\bar{\gamma}_i$ to be given by a volume-independent value, tending to $1$ as $W\rightarrow \infty$. For each disorder realization, we use 50...100 eigenstates $|n\rangle$ and for each set of model parameters at least 2000 disorder realizations of the fields $h_i$ and bubble Hamiltonian $\hat{R}_l$.  


\begin{figure}[h]
	\centering
	\includegraphics{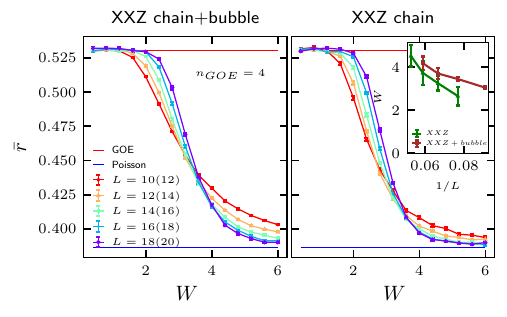}
	\caption{Average gap ratio \cite{oganesyan_localization_2007} $\overline{r}$ for the disordered field XXZ spin chain of size $L$ coupled to a thermal bubble of size $n_\text{GOE}=4$ (left) and the same system \emph{without} bubble and chain length $L+2$ (right). Inset: Pair wise crossing of gap ratio $\overline{r}(W)$ as function of $1/L$ for both systems.}
	\label{fig:gap_ratio} 
\end{figure}

\section{Shift of MBL transition in the interacting chain}
In an infinitely long disordered chain, it should not matter whether we add a thermal bubble or not, because thermal Griffiths regions acting as ergodic patches are expected to be present anyhow. 
In a chain of moderate size, we expect that a substantial fraction of samples appears to be localized simply by lack of ergodic regions. 
This would lead to a shift of the apparent critical disorder value $W_c$ for short chains if one compares the isolated chain to our model where we add a bubble by hand. In Fig.~\ref{fig:gap_ratio}, we compare the disorder averaged gap ratio $\overline{r}$ of adjacent energy gaps in the middle of the spectrum given  of the isolated XXZ chain (right) to XXZ chain-bubble system (left) as a function of disorder strength $W$ for different system sizes $L$. Using the crossing of  $\overline{r}$ of size $L$ and $L+2$ as a proxy for the apparent critical point at this length scale, the inset shows that it is indeed the case that the crossings appear at slightly larger disorder strengths in the presence of the bubble.
However, due to large statistical uncertainty and significant finite size effects, this crossing analysis is not a reliable way to pin down the critical point.

\section{Change of local thermality in interacting chains}



\begin{figure}[h]
	\centering
	\includegraphics{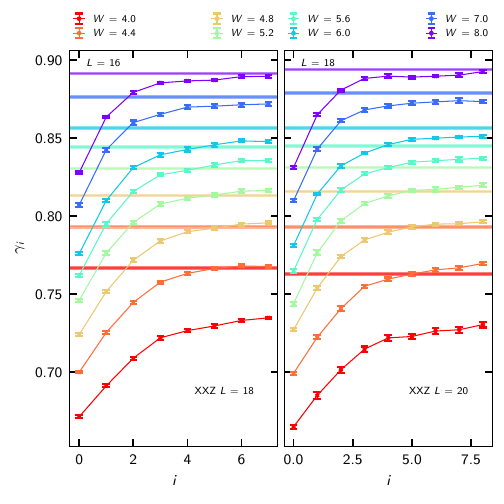}
	\caption{Purity $\gamma_i$ as function of the distance from the bubble for different disorder strengths $W$. Left (Right) panel shows system size $L=16$ ($L=18$) and $n_\text{GOE}=4$. Horizontal shaded areas correspond to the $\gamma_{\text{XXZ}}$ in the absence of the bubble with $L+2$ chain length compared to the bubble-chain system.}  
	\label{fig:purity_mean}
\end{figure}
\begin{figure}[h]
	\centering
	\includegraphics{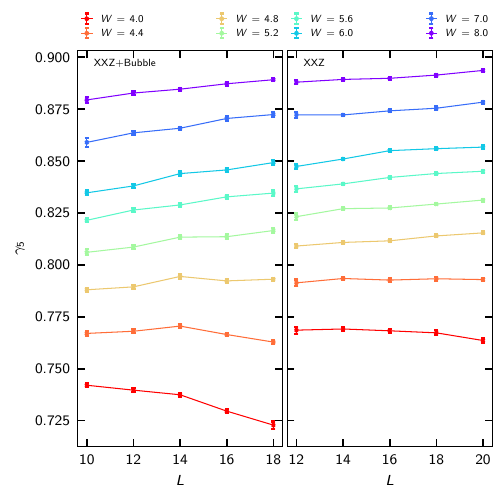}
	\caption{Purity $\gamma_i$ at fixed distance $i=5$ from the bubble plotted as function of system size $L$ for different disorder strengths $W$. Left panel: XXZ spin chain of length $L$ with bubble of size $n_{\text{GOE}}=4$. Right panel: XXZ Heisenberg chain of size $L$. Average is taken over disorder realizations and eigenstates. } 
	\label{fig:purity_mean_L}
\end{figure}
\begin{figure}[h]
	\centering
	\includegraphics{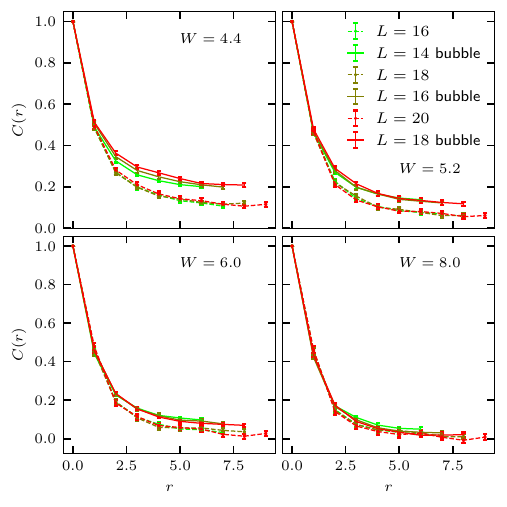}
	\caption{Average purity correlation $C(r)$ as function of the distance from the bubble for disorder strengths $W=4.4,5.2,6.0,8.0$. Dashed lines correspond to bubble-less system of size $L+2$ while solid lines to bubble-full system of size $L$}    
	\label{fig:purity_correlation}
\end{figure}
%


We investigate the one-site purities $\gamma_i$ as a measure of local thermality. Throughout this paper, we consistently use the labels ``localized" and ``thermal" as inferred from the crossing point of $\overline{r}$ (see Fig.~\ref{fig:gap_ratio}), i.e.\ with the critical point at $W\approx3.5-4$, even though recent works locate the transition at larger $W$. 
Inspecting the average purities in the presence and absence of the thermal bubble (see Fig.~\ref{fig:purity_mean}) is clear that the effect of the bubble at long distance is rather small and subtle. 
Therefore, it is useful to first map out what the theory of quantum avalanches predicts. 

\noindent \textbf{Theoretical background}   For a simple avalanche model of our setup, we assume the existence of an \emph{apparent} localization length $\xi$, describing the system in absence of the bubble, and a number $p$ giving the probability that the bubble can kickstart a thermalization process in its near vicinity, see Ref.~\cite{crowley_avalanche_2020} for a detailed discussion and arguments why $p \ll 1$ in our setup. A simple model, based on an unrealistic dichotomy between ergodicity and localization, leads (see SM) to the following very rough prediction: 
\begin{equation}\label{eq:toy_equation}
(\bar\gamma_i-1/2) \approx (1-p)(1-e^{-|i|/\xi})(\bar\gamma_{\mathrm{XXZ}-1/2}) + pd^{-1/2}_{\text{eff},i}
\end{equation}
with $\bar \gamma_i$ the average purity at site $i$ and $\bar\gamma_{\mathrm{XXZ}}$ the average purity in the bubble-less system. For $p=0$, we recover a perfectly localized system, whereas for $p=1$, the system is ergodic but the average purity still increases with $i$, because  the effective dimension $d_{\text{eff},i}$ depends on the distance to the bubble. This effective dimension 
is defined as
$
d_{\text{eff},i}= e^{-2i/\xi} d_{\mathrm{therm}}
$
with $d_{\mathrm{therm}}$ the total Hilbert space dimension of the thermal region, given by 
$d_{\mathrm{therm}}= n_\text{GOE}^{\frac{1}{1-\xi/\xi_*}}$ if $\xi<\xi_*$ and $d_{\mathrm{therm}}=n_\text{GOE}2^L$ if $\xi>\xi_*$, where $\xi_*=1/\log 2$ is the critical localization length. 
According to the above formula Eq.~\eqref{eq:toy_equation}, there are two regimes in which $\bar\gamma_{\mathrm{XXZ}}-\bar \gamma_i$ does not decay to zero, or only very slowly. \begin{itemize}
\item[I)] $\xi> \xi_*$. Here, the bubble thermalizes a fraction $p$ of samples, resulting in a shift $\bar\gamma_{\mathrm{XXZ}}-\bar \gamma_i$ which remains finite as $i\to\infty$, even though it decreases due to the decrease of $d_{\text{eff},i}$.
\item[II)] $\xi$ approaches $\xi_*$ from below. Then $\bar\gamma_{\mathrm{XXZ}}-\bar \gamma_i \to 0$ at large $i$, but the decay is arbitrarily slow when  $\xi \to \xi_*$, because the decay of $d_{\text{eff},i}$ is arbitrarily slow.
\end{itemize}
In practice,  I) and II) are of course hard to distinguish.\\
\noindent \textbf{Large distance behaviour} In Fig.~\ref{fig:purity_mean}, we indeed see a sign of influence of the bubble that does not, or only very slowly, decay with distance from the bubble.  We observe that the purity in the presence of the bubble seems to tend to an asymptotic value that is significantly lower than the value for the bubble-less system. This seems to be the case up to disorder strength $W=5.6-6$, after which the signal, i.e.\ the difference $\bar\gamma_{\mathrm{XXZ}}-\bar \gamma_i$, becomes comparable to the error bars. 
We will analyze this further below, but let us first note that it is far from clear whether our system sizes are large enough to speculate about the limit $i\to\infty$. Indeed, in Fig.~\ref{fig:purity_mean_L}, we investigate the dependence of the purity at fixed distance on increasing system size $L$. We see that for disorder values at and above $W=5.2$, the purity $\gamma_5$ is actually increasing with increasing system size. This is an effect that is not even accounted for in our model equation and it should be interpreted as a sign that finite-size effects are still important (see SM material Sec.~\ref{sec:additional_data} for further analysis). 
Another way to look for thermalization induced by the bubble is through correlations between different sites of the chain. In that spirit we look at statistical correlations between two purities $\gamma_i$ and $\gamma_j$. Those are defined as 
\begin{equation}\label{eq:purity_corr}
C(r) = \dfrac{\overline{(\gamma_{0}-\bar\gamma_{0})(\gamma_{r}-\bar\gamma_{r})}}{\sigma_{0}\sigma_{r}},
\end{equation}
where $\gamma_{i}-\bar\gamma_{i}$ captures the purity fluctuations around its mean and $\sigma^2_i = \overline{(\gamma_{i}-\overline{\gamma_i})^2}$ is the variance of the purity at site $i$. The quantity $C(r)$ is the Pearson correlation coefficient between purities at site next to the bubble ($i=0$) and sites at distance $r$ from that spin with $r=0,1,...,L/2$. This quantity for different disorder and system size is shown in Fig. \ref{fig:purity_correlation}. Interestingly, for $W=4.4$ the bubble seems to be enhancing correlations compare to the bubble-less case. This effect is even growing with system size and it persists for $W=5.2$ and, up to some extent, for $W=6.0$. 

Since the signals that we observe, the difference $\bar\gamma_{\mathrm{XXZ}}-\bar\gamma_i$ and the correlations in Fig.~\ref{fig:purity_correlation}, are rather small, there is the concern that they might be caused by some artefact.  
For example, even in a well-localized system, in the absence of any avalanches, the purity-purity correlation $C(r)$ from Eq.~\eqref{eq:purity_corr} has a non-zero limit as $r\to\infty$ of order $\frac{C}{L^2}$ with $C \propto \frac{1}{W^2}$ in the high-disorder limit $W\to\infty$, see Appendix for more details. While we do not see any clear mechanism how this might pollute our analysis, we find it hard to exclude it. 

\section{Conclusion}

We have studied the effect of ergodic inclusions modeled by local random matrices in disordered spin chains of up to 20 sites. We report little or no drift of the thermal-to-localized transition in the average level spacing compare to the bubble less-case, which was also observed in similar settings \cite{goihl_exploration_2019}. We also investigate long distance effects of the bubble by looking at functions of the local magnetization expectation value, in this case the purity and its fluctuations. There, we have numerically witnessed a potentially divergent, long wavelength, response of an apparently localized chains (as estimated by ED studies like \cite{luitz_many-body_2015}) to an ergodic bubble. The effect is weak and not unambiguous, but it is compatible with the avalanche theory proposed in earlier works. Importantly, such weak correlations may be influenced by spurious long-range correlations due to the $U(1)$ symmetry (see Sec.~\ref{sec;purity_correlations} in SM), although we can not confirm this from our numerics.

Recent studies also point out weak but persisting correlations produced by many-body resonances in the same disorder regime we study \cite{morningstar_avalanches_2022,crowley_constructive_2022}. Our results suggest that the bubble is effectively enhancing such weak correlations when the system looks well localized in average. However it is yet not clear how to directly relate the observed signal with the many-body resonances.

\begin{acknowledgments}
L.C. gratefully acknowledges funding by the U.S. ARO Grant No. W911NF-21-1-0007. All statements of fact, opinion or conclusions contained herein are those of the authors and should not be construed as representing the official views or policies of the US Government. This project was supported by the Deutsche Forschungsgemeinschaft (DFG) through the cluster of excellence ML4Q (EXC 2004, project-id 390534769). We further acknowledge support from the QuantERA II Programme that has received funding from the European Union’s Horizon 2020 research innovation programme (GA 101017733), and from the Deutsche Forschungsgemeinschaft through the project DQUANT (project-id 499347025).
\end{acknowledgments}

\bibliography{bubble-MBL}
\newpage
\clearpage

\section{Supplementary material}

\subsection{Parameters of the model}\label{sec:beta}

\begin{figure}[h]
	\centering
	\includegraphics{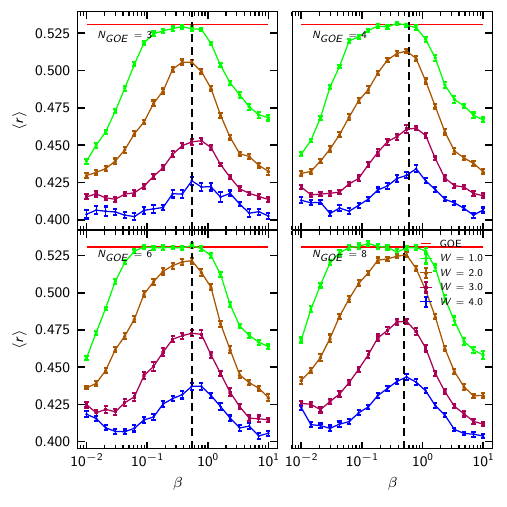}
	\caption{Average gap ratio $\langle r\rangle$ as function of random matrix band width $\beta$ for different $n_\text{GOE}={3,4,6,8}$. The black dashed line stands for $\beta = 0.55,0.6,0.55,0.5 $ respectively the chosen values in our simulations that cut off the plateus where the spin chain-bubble mixing is maximal for the relevant parameters.}  
	\label{fig:beta}
\end{figure}

In this model two systems, the spin chain and the bubble are coupled via $H_{\mathrm{coupling}}$ given by Eq.~\eqref{eq:coupling}, where Eq.~\eqref{eq:randommat} shows the definition of the random matrices that described the bubble Hamiltonian. For coupling efficiently both systems their spectra should have comparable band widths which allows maximum possible mixing of their eigenstates. The parameter $\beta$ controls the average band width of the random matrices, then an optimal choice of $\beta$ enhances the mixing of the two systems and therefore maximizes the probabilities of the avalanche to happen. In order to do so, the average gap ratio $\langle r \rangle$ is computed for different values of band width $\beta$ at several disorder strength (Fig. \ref{fig:beta}). We see that $\langle r \rangle$ always has maxima and they happen to be almost at the same $\beta$. Choosing $\beta$ such that the average gap ratio is always maximum ensures that the spin chain and the bubble are strongly coupled. For $1.0 < W < 4.0 $ and $n_\text{GOE} =3,4,6,8 $ we found that $\beta = 0.55,0.6,0.55,0.5 $ respectively satisfy this requirement. It could be that at $W>4.0$ the band width we chose is de-tuned from the optimal value by a bit, however we do not expect this to change the main results since the maxima blunt and the curves become flatter when disorder is increased.   

\subsection{Local magnetization}\label{sec:local_magnetization}

\begin{figure}
	\centering
	\includegraphics{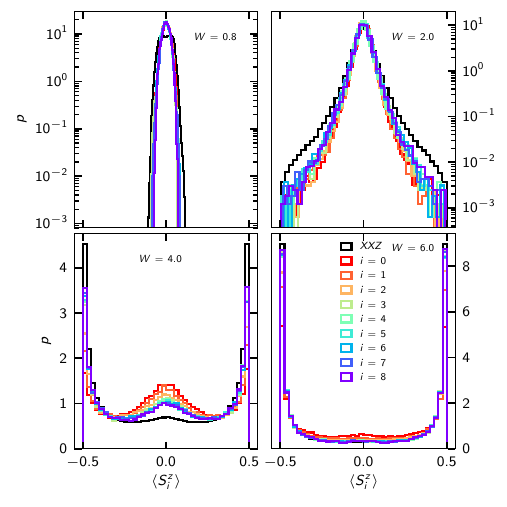}
	\caption{Distribution of local magnetization for $\mathrm{XXZ}$ chain with (color lines) and without bubble (black lines) for system size $L=20$ and $L=18$ respectively (bubble size $n_{\text{GOE}}=4$). Each color denotes the spin at site $i$ from the bubble (see Fig.~\ref{fig:setup} for exact layout) }  
	\label{fig:sz_distribution}
\end{figure}

In this section we present the distribution of local magnetization in the system with and without bubble (see Fig.~\ref{fig:sz_distribution}). There we confirm that at low disorder the bubble has no effect beyond a small shift to ``more Guassian" distributions. At disorder $W=4.0$, which is supposed to be in the critical region, we see that the localized peaks at $\langle S^z\rangle=\pm 1$ loose weight in favor of the a central ``thermal" peak in the presence of the bubble. The latest suggest that an important fraction of samples thermalizes in the presence of the bubble. Unfortunately we could not quantify the fraction of thermalizing samples in a more rigorous manner.
A similar but more detailed analysis of distributions was done in Ref.\cite{luitz_how_2017} where non-interacting localize bits were coupled to a ergodic bubble. Unfortunately finite size effects are stronger in our current setup, thus avoiding trustworthy conclusions from such distributions. 

\subsection{Additional data}\label{sec:additional_data}

\begin{figure}
	\centering
	\includegraphics{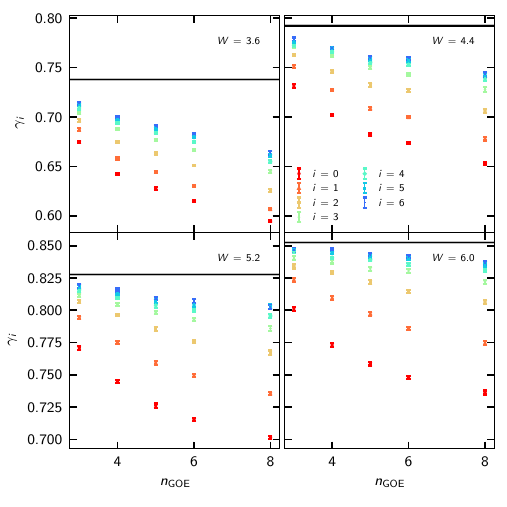}
	\caption{Purity $\gamma_i$ average over disorder realization and eigenstates as function of bubble size $n_\text{GOE}$, several disorder strengths $W$ and different distances $i$ from the bubble. The black line corresponds to $\gamma_\text{XXZ}$ in absence of the bubble ($L=14$). The chain length $L=14$ is fixed. $\beta =\{ 0.55,0.6,0.58,0.55,0.5\}$ is tuned to be the optimal value for each bubble size $n_\text{GOE}=\{3,4,5,6,8\}$}. 
	\label{fig:std_Sz_grain}
\end{figure} 

\begin{figure}
	\centering
	\includegraphics{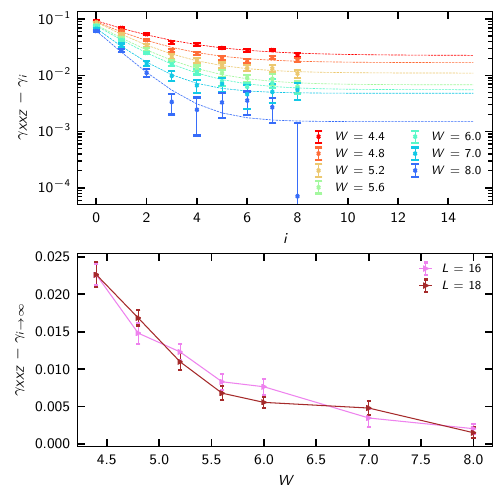}
	\caption{Upper panel: difference $\overline{\gamma}_{XXZ}-\overline{\gamma}_i$ for different disorder strength, $L=18$ and $n_\text{GOE}=4$. Dashed lines are fittings $\gamma_0+Ae^{-i/\xi}$, with $\gamma_0$ as the limiting value $i\rightarrow\infty$ . Lower panel: $\gamma_0$ as function of $W$ for system sizes $L=16,18$. In both panels, $\overline{\gamma}_{XXZ}$ is computed using $L+2$ chain sites compared to the shown system sizes.}  
	\label{fig:delta_diff}
\end{figure}


In this section we present a further analysis on the purity $\gamma_i$ on localized $\mathrm{XXZ}$ chains with and without bubble. Although the system sizes are small and quantitative results are not conclusive, we find instructive to sketch the qualitative signals of thermalization due to the presence of the bubble. 

In Fig.\ \ref{fig:std_Sz_grain}, we study the limit where the bubble size $n_\text{GOE}$ growth while the length of the chain is kept fixed. The most visible feature in these plots is that the purity \ $\bar \gamma_i$ decreases when a bubble is coupled to the chain, with bigger bubbles having a larger effect. This trend is obvious and expected, because coupling an ergodic bubble can, on average, only increase the thermality. Hence, this trend does not give us any non-trivial information on the nature of the ergodic-to-MBL transition.  

To go beyond this, we note first that at very large distance $i\to\infty$, one expects that $\bar{\gamma}_i- \overline{\gamma}_{\mathrm{XXZ}} \to 0$ independently of whether the systems is localized or ergodic, simply because in a very large system, arbitrarily large bubbles will occur naturally.
Numerically, at disorder strengths $W<7$, we observe however that the difference $\overline{\gamma}_{XXZ}-\bar{\gamma}_i $ seems to decay very slowly, if at all.  In Fig. \ref{fig:delta_diff}, we remark that this function is actually reasonably fitted by an exponential decay law added to a nonzero constant $\delta_0$.  For $W\geq 7$, the deviation of $\delta_0$ from $0$ is no longer statistically meaningful. 
It appears hence that for disorder strengths $W<7$, there is a effect of the bubble at large distances (see discussion in the main text). This effect is weak, for example $5\%$ at $W=5.2$, but; however due to the smallness of the systems considered we can not rule out attributing this signal to finite size effects. 


\subsection{Quantum avalanche theory}
Let us recall the avalanche theory \cite{de_roeck_stability_2017} for a disordered spin chain coupled to a thermal inclusion. If the chain is isolated from the bubble, it is assumed to be many-body localized, forming local integrals of motion (LIOM) \cite{serbyn_local_2013,huse_phenomenology_2014}. We assume that the LIOMs can be described by a localization length $\xi$, whose meaning is that a LIOM centered at some site of the chain is coupled with strength $e^{-\ell/\xi}$  (operator norm of the coupling) to sites at distance $\ell$. 
The ergodic bubble is assumed to be perfectly thermal and is described by a Hilbert space of dimension $n_\text{GOE}$.

The logic of the theory is based on an iterative application of perturbation theory: We start with the ergodic bubble and couple in the first step only the LIOMs  which have the largest overlap (and therefore coupling) with the bubble. Those are the LIOMs centered at the sites adjacent to the ergodic bubble. There are two of those sites because we consider a periodic chain. If the coupling is non-resonant, we declare that the adjacent LIOMs have not been thermalized, and the analysis stops there. If the coupling is resonant, then we declare the adjacent spins thermalized and we imagine that the ergodic region has now dimension $4n_\text{GOE}$ instead of $n_\text{GOE}$, i.e.\ the LIOMs have been \emph{absorbed} by the ergodic bubble. 
This is now repeated, proceeding further and further away from the original bubble, until either the adjacent LIOMs are non-resonant, or until all LIOMs have been thermalized. 

Let us make this quantitative. We
assume that the bubble has thermalized all LIOM's within a distance $\ell$, leading to a delocalized (ergodic) space with dimension $d_{\mathrm{therm}}=n_\text{GOE} 4^{\ell}$. We will now only keep the overall scaling of the relevant terms to obtain a criterion for thermalization:
We couple the next LIOM with strength of order $ e^{-\ell/\xi}$. We use random matrix theory to compute a typical matrix element of such a coupling and  the typical level spacing is estimated simply as $\propto d_T^{-1}$. We then get for the quantity distinghuishing resonance from non-resonance:
\begin{equation}\label{eq: condition pt}
\frac{\text{matrix element}}{\text{level spacing}}  \sim   e^{-\ell/\xi} d_{\mathrm{therm}}^{1/2}  =  e^{-\ell/\xi}  (d_G 4^{\ell})^{1/2}
\end{equation}

The thermalization proceeds until this ratio is of order $1$, hence up to $\ell=\ell_c$ given by
$$
\ell_c=  (1/\xi-1/\xi_*)^{-1} \tfrac12 \log n_\text{GOE},\qquad  \xi^{-1}_*=\log 2,
$$
where $\xi_*$ is the critical localization length. This formula shows that $\ell_c$ is finite if $\xi<\xi_*$ and infinite otherwise. Due to the final length of the system, the correct distance up to which LIOMS are thermalized is of course the minimum of $\ell_c$ and $L/2$.
The Hilbert space dimension of the thermal region is then, for  $\xi<\xi_*$, 
$$
d_{\mathrm{therm}}=  n_\text{GOE}4^{\ell_c}=  n_\text{GOE}^{\frac{1}{1-\xi/\xi_c}}
$$
Taking also here into account the finite size, we arrive at the expression given in the main text. 

Let us now briefly comment on the \emph{effective dimension} which was mentioned in the main text. This concept was introduced in \cite{luitz_how_2017} as follows. The spectral form factor of local observables at site $i$ have a width that is way smaller than of the order of local energy scales, as it is in a normal many-body system. Instead, this width is of order of the mean level spacing times the effective dimension $d_{\text{eff},i}=e^{2i/\xi}d_{\mathrm{therm}}$. This formula was also verified numerically in \cite{luitz_how_2017}.
It is the smallness of this width, or, equivalently, the fact that $d_{\text{eff},i}$ is way smaller than the total Hilbert space dimension, that gives rise to the extremely slow dynamics in this system, as one goes far away from the ergodic bubble. 

Finally, we return to equation \eqref{eq:toy_equation}, which can be viewed as a consequence of the above discussion.
In the case (occuring with probability $1-p$) the ergodic bubble did not kick-start an avalanche, i.e.\ when the coupling to the first LIOM was too weak to thermalize it, we naively approximate the influence of the ergodic bubble on the purity as 
$(\bar \gamma_i-1/2)\approx (1-e^{-|i|/\xi})  (\bar\gamma_{\mathrm{XXZ}-1/2})$. This simply reflects the fact that the purity at site $i$ is influenced by the delocalized states in the ergodic bubble when $|i|/\xi$ is not too large.
 If an avalanche was indeed started, we can use the theory described above. The purity is computed from local observables and hence it can be computed assuming that one has an ergodic system with dimension $d_{\text{eff},i}$. Some simple random matrix theory leads then to the conclusion that this purity is given by $d_{\text{eff},i}^{-1/2}$. 





\subsection{Purity correlations}\label{sec;purity_correlations}

We compute the purity perturbatively in the high-disorder limit. This will confirm the existence of a residual purity-purity correlation at large distance.

\subsubsection{Perturbative expression for purity} \label{sec:perturbative}

We focus on the standard MBL model
\begin{eqnarray}
H=\sum_i h_i \sigma^z_i + J(\sigma^+_i  \sigma^-_{i+1} +hc)  + J_z \sigma^z_i  \sigma^z_{i+1}.
\end{eqnarray}
Let $\eta$ be classical configurations $\eta_i =\pm 1$ and let $h_i \in [-W,W]$. We assume that we are in the MBL phase and the eigenstates connected to the classical configurations.
We will actually do a perturbative expansion in $J/W$ (locator expansion) that disregards resonant regions. It yields an expression for the eigenstates given by 
\begin{eqnarray}
|\Psi(\eta)\rangle=  e^{iA}|\eta\rangle,
\end{eqnarray}
where $A=\sum_{i} A_{i,i+1}$ is a sum of local Hermitian terms, determined by the condition
\begin{eqnarray}
\langle \eta| A_{i,i+1} | \eta'\rangle=  \frac{1}{E^{(0)}(\eta')-E^{(0)}(\eta) } \langle \eta|   J(\sigma^+_i  \sigma^{-}_{i+1} +h.c.) | \eta'\rangle
\end{eqnarray}

Depending on $\eta$, there is either zero or one $\eta'$ such that this matrix element is non-zero. We compute the unique non-zero matrix element as
\begin{eqnarray}
\langle \eta| A_{i,i+1} | \eta'\rangle=  \frac{J(\eta_{i+1}-\eta_i)}{4(h_{i+1}-h_i)+2J_z(\eta_{i+2}-\eta_{i-1})} 
\end{eqnarray}
Now we compute
\begin{eqnarray}
\langle \Psi(\eta)| \sigma_i^z   |\Psi(\eta)\rangle & = &  
\langle \eta| \left(\sigma_i^z +i [A,\sigma_i^z] - \tfrac{1}{2}  [A,[A,\sigma_i^z]]  \right) 
|\eta \rangle  \\  
&& + O((J/W)^3). 
\end{eqnarray}
It is easy to see that the first order term $ [A,\sigma_i^z]$ does not contribute to the expectation value so we calculate the second order term as well, which will be of order $(J/W)^2$. This suggests that we should have considered also the second order in the expression for $A$, but this is fortunately not the case. This is  because the term linear in the second order $A$ will again be a commutator $[A^{(2)},\sigma^z_i]$, whose expectation value will vanish, just as it did for the first order $A$. 
The result is hence
\begin{align}
Z_i:=-\langle \eta| \left( \tfrac{1}{2}  [A,[A,\sigma_i^z]]  + \ldots  \right)   |\eta \rangle &=  \sum_{\eta'}  (\eta'_i-\eta_i) | \langle \eta| A | \eta'\rangle|^2 
\end{align}
which can be computed to give
\begin{eqnarray}
Z_i & = & -2\eta_i \left[\frac{J(\eta_i-\eta_{i+1})}{h_{i+1}-h_i +2J_z(\eta_{i+2}-\eta_{i-1})} \right]^2 + \nonumber \\
&& -2\eta_i \left[\frac{J(\eta_i-\eta_{i-1})}{h_{i}-h_{i-1} +2J_z(\eta_{i+1}-\eta_{i-2})} \right]^2 
\end{eqnarray}
We have then
\begin{eqnarray}
\langle \Psi(\eta)| \sigma_i^z   |\Psi(\eta)\rangle \approx \eta_i+Z_i
\end{eqnarray}

Note that $Z_i$ is such that the left-hand side is indeed in $[-1,1]$, as it should.

Now from this we can compute the purity
\begin{eqnarray}
P_i(\eta)= |\langle \Psi(\eta)| \sigma_i^z   |\Psi(\eta)\rangle|^2 \approx 1+2\eta_i Z_i +Z_i^2 
\end{eqnarray}
To order $(J/W)^2$, we have then 
\begin{eqnarray}
P_i(\eta) & = & 1-2 \left[\frac{J(\eta_i-\eta_{i+1})}{h_{i+1}-h_i +2J_z(\eta_{i+2}-\eta_{i-1})} \right]^2 + \nonumber \\ && -2\left[\frac{J(\eta_i-\eta_{i-1})}{h_{i}-h_{i-1} +2J_z(\eta_{i+1}-\eta_{i-2})} \right]^2
\end{eqnarray}

This is a well-function of $\eta$ and this ends the calculation of the purity.  

\subsubsection{Correlation of purity}

Let us now for simplicity disregard the interaction, setting $J_z=0$ and use the fact that 
\begin{eqnarray}
(\eta_i-\eta_{i+1})^2=  2(1 -\eta_i\eta_{i+1})
\end{eqnarray}
Then 
\begin{eqnarray}
P_i(\eta) & = & 1- ( \tfrac{2J}{(h_{i+1}-h_i)})^2  (1 -\eta_i\eta_{i+1})
- \nonumber \\
&& ( \tfrac{2J}{(h_{i-1}-h_i)})^2 (1 -\eta_i\eta_{i-1})
\end{eqnarray}

Now we are ready to calculate 
\begin{eqnarray}
\langle P_i(\eta) P_j(\eta) \rangle -  \langle P_i(\eta)\rangle  \langle P_j(\eta)\rangle 
\end{eqnarray}
when $i,j$ are far apart.  This correlation will be of the form
\begin{eqnarray}
C \frac{J^2}{W^2}  \left( \langle \eta_i\eta_{i+1}  \eta_j\eta_{j+1}  \rangle  -  \langle \eta_i\eta_{i+1} \rangle \langle   \eta_j\eta_{j+1}  \rangle \right)
\end{eqnarray}
where we replaced the disorders by some typical value.
The expression between brackets is now computed explicitly, yielding the value
$2/L^2$.




\end{document}